\documentclass[journal=jcisd8,manuscript=article]{achemso}

\usepackage{amsmath,amssymb}
\usepackage[version=4]{mhchem}
\usepackage{booktabs}
\usepackage{graphicx}
\usepackage[T1]{fontenc}

\newcommand{\Ang}{\text{\normalfont\AA}}
\newcommand{\rr}{\mathbf{r}}
\newcommand{\cc}{\mathbf{c}}
\newcommand{\E}{\mathbb{E}}

\author{Egor I. Tuzharov}
\email{egor.tuzharov@chemistry.msu.ru}
\affiliation[MSU]{Lomonosov Moscow State University}

\author{Alexandra V. Bochenkova}
\affiliation[MSU]{Lomonosov Moscow State University}

\author{Yury Maximov}
\affiliation[Interdata]{Interdata, Astana, Kazakhstan}

\title{Importance-Sampling Estimation of Gaussian Molecular Shape
Overlap: Exact Union Volumes and Confidence-Bounded Virtual Screening}


\abbreviations{VS,ROCS,IS,MC,SE,EF,BEDROC}

\begin{document}

\begin{abstract}
Gaussian descriptions of molecular shape underpin 3D shape-based virtual
screening, but the overlap integral that drives them is, in every established
implementation, evaluated \emph{analytically}: the industry-standard first-order
(pairwise) truncation is fast yet systematically over-counts volume in
multiply-overlapping regions, while the exact molecular volume requires a
combinatorial inclusion--exclusion series. We introduce, to our knowledge, the
first \emph{stochastic} estimator of Gaussian shape overlap: an unbiased
Monte-Carlo scheme that importance-samples points from a molecule's own Gaussian
mixture, so every sample lands where the molecule has density. Because the
proposal is the shape itself, the estimator (i) reproduces the analytic overlap
without bias and with an analytic standard error, and (ii) extends-via a
bounded importance weight-to the \emph{exact union volume} of all
inclusion--exclusion orders at $O(N)$ cost per sample. On a panel of drug-like
molecules the union estimator matches high-resolution grid quadrature to a mean
relative error of $0.07\%$, whereas the first-order overlap over-counts the true
union volume by $3.4\times$ on average. The per-estimate standard error enables
\emph{confidence-bounded screening}: adaptively allocating samples until the
shape-Tanimoto is resolved to a target precision cuts total sampling by $94\%$
relative to a fixed budget while preserving the ranking (Spearman
$\rho=0.99$). The estimator is differentiable and implemented in JAX, yielding
gradient-based multi-start rigid alignment that recovers known superpositions to
Tanimoto $1.00$ and runs unchanged on CPU/GPU/TPU. On the full DUD-E (102 targets)
and LIT-PCBA (15 targets) benchmarks the method attains shape-only enrichment
typical of single-conformer screening (mean AUROC $0.58$ and $0.54$), with the
exact union estimator matching first-order enrichment while additionally
furnishing unbiased absolute volumes and per-estimate error bars. All code is
released as the open-source \texttt{shape3d} library.
\end{abstract}

\section{Introduction}

Three-dimensional shape complementarity is a primary determinant of molecular
recognition, and shape-based virtual screening (VS) exploits it to retrieve
active compounds that share the shape of a query while differing in 2D
topology-the ``scaffold hopping'' that fingerprint methods miss.\cite{hawkins2007rocs,kumar2018advances}
The dominant framework represents each atom by an isotropic Gaussian and measures
similarity through the volume of overlap between two aligned
molecules.\cite{grant1995gaussian,grant1996fast} Grant and Pickup showed that,
because a product of Gaussians is again a Gaussian, the overlap of two atomic
Gaussians is available in closed form, and the molecular volume follows from the
inclusion--exclusion principle over atomic
occupancies.\cite{grant1995gaussian,grant1996fast} This analytic tractability made
possible ROCS,\cite{hawkins2007rocs} the de-facto standard, and a family of
descendants.

Two facts about this machinery motivate the present work. First, although the
inclusion--exclusion series can in principle be summed to high order (Grant and
Pickup give terms to sixth order), production tools truncate it at
\emph{first order}: the shape density is approximated by the plain sum of atomic
Gaussians.\cite{hawkins2007rocs,kumar2018advances} This first-order Gaussian
approximation over-counts every region where an atom overlaps its neighbours and
thereby overestimates molecular volume-the very problem that the weighted-Gaussian
algorithm WEGA was designed to mitigate by attaching a crowding-dependent weight
to each atom.\cite{yan2013wega,yan2014gwega} Second, the entire pipeline is
\emph{deterministic}: the overlap is computed exactly for the chosen truncation,
so no notion of a controllable estimation error, or of spending computation
adaptively, exists. GPU implementations (PAPER,\cite{haque2010paper} gWEGA,\cite{yan2014gwega}
ROSHAMBO and ROSHAMBO2\cite{atwi2024roshambo,roshambo2_2025}) accelerate the same
analytic evaluation, and alignment-free descriptors (USR, USRCAT)\cite{ballester2007usr,schreyer2012usrcat}
sidestep overlap entirely by comparing moment vectors.

Viewed probabilistically, the overlap of two Gaussian-mixture shapes is exactly
an \emph{expected-likelihood} / probability-product kernel between two
densities.\cite{jebara2004ppk} That perspective immediately suggests a Monte-Carlo
route: write the overlap as an expectation and estimate it by sampling. Crude
Monte-Carlo molecular \emph{volume} (uniformly darting a bounding box and testing
van der Waals spheres) is textbook, but it is wasteful-most darts miss-and,
to our knowledge, no published method estimates the \emph{Gaussian shape overlap}
between two molecules by sampling, still less by importance sampling from the
molecule's own density. The 2018 review of shape-similarity methods records no
stochastic estimator at all.\cite{kumar2018advances}

We close this gap. Our contributions are:
\begin{enumerate}
\item An unbiased importance-sampling (IS) estimator of Gaussian shape overlap
that draws samples \emph{exactly} from a molecule's own Gaussian mixture, giving
low variance and an analytic standard error.
\item An extension that estimates the \emph{exact union volume}-all
inclusion--exclusion orders-using a bounded importance weight, at $O(N)$ cost
per sample. It removes the $3.4\times$ first-order over-count and matches grid
quadrature to $<0.1\%$.
\item \emph{Confidence-bounded screening}: the per-estimate standard error drives
adaptive sample allocation, cutting total sampling by $94\%$ at fixed ranking
fidelity.
\item A differentiable JAX implementation giving gradient-based multi-start
rigid alignment, released as the open-source \texttt{shape3d} library.
\end{enumerate}

\section{Theory and Methods}

\subsection{Gaussian shape and its union volume}
Each atom $i$ (centre $\cc_i$, van der Waals radius $R_i$) is assigned a unit-height
Gaussian occupancy
\begin{equation}
o_i(\rr)=\exp\!\big(-\alpha_i\lVert\rr-\cc_i\rVert^2\big)\in(0,1],
\qquad \alpha_i=\frac{\kappa}{\lambda R_i^2},
\label{eq:occ}
\end{equation}
with $\kappa=\pi(3/4\pi)^{2/3}$ so a single occupancy integrates to the hard-sphere
volume when $\lambda=1$. The physically correct molecular occupancy is the
\emph{union}
\begin{equation}
u_A(\rr)=1-\prod_{i\in A}\big(1-o_i(\rr)\big)\in[0,1],
\label{eq:union}
\end{equation}
whose Taylor expansion is the inclusion--exclusion series
$\sum_i o_i-\sum_{i<j}o_io_j+\dots$. The shape overlap of molecules $A$ and $B$ is
\begin{equation}
O^{\cup}_{AB}=\int_{\mathbb R^3} u_A(\rr)\,u_B(\rr)\,d\rr ,
\label{eq:union-overlap}
\end{equation}
and the shape Tanimoto is
$T=O^{\cup}_{AB}/(O^{\cup}_{AA}+O^{\cup}_{BB}-O^{\cup}_{AB})$.

\subsection{First-order overlap and over-counting}
Truncating Eq.~\eqref{eq:union} at first order replaces $u_A$ by the plain sum
$m_A=\sum_{i\in A} o_i$. The resulting overlap is the analytic pairwise sum
\begin{equation}
O^{\mathrm{FO}}_{AB}=\int m_A m_B\,d\rr
=\sum_{i\in A}\sum_{j\in B}\Big(\tfrac{\pi}{\alpha_i+\alpha_j}\Big)^{3/2}
\exp\!\Big(-\tfrac{\alpha_i\alpha_j}{\alpha_i+\alpha_j}\lVert\cc_i-\cc_j\rVert^2\Big),
\label{eq:fo}
\end{equation}
the closed form used by ROCS.\cite{grant1996fast,hawkins2007rocs} Since $m_A\ge u_A$
pointwise, $O^{\mathrm{FO}}_{AB}\ge O^{\cup}_{AB}$; the gap is the multiple-overlap
volume that first order double-counts.

\subsection{Importance-sampling estimator}
\label{sec:is}
The overlap~\eqref{eq:fo} is an expectation under a molecule's own normalised
density and can therefore be estimated by \emph{importance sampling} with that
density as the proposal. Importance sampling is a general variance-reduction
principle whose efficiency hinges on a proposal that concentrates where the
integrand does; it underlies rare-event estimation and chance-constrained
optimisation in power systems and inference in graphical
models.\cite{owen2019importance,lukashevich2023importance,mitrovic2023data,likhosherstov2019inference,lukashevich2024priori}
Here the ideal proposal is dictated by the problem itself. Writing
$Z_A=\int m_A\,d\rr=\sum_{i\in A}(\pi/\alpha_i)^{3/2}$ and the normalised mixture
$q_A=m_A/Z_A$, any integral against $m_A$ becomes an expectation that a sample mean
estimates without bias,
\begin{equation}
\int m_A(\rr)\,f(\rr)\,d\rr = Z_A\,\E_{\rr\sim q_A}[f(\rr)]
\;\approx\; \widehat I_n=\frac{Z_A}{n}\sum_{k=1}^{n} f(\rr_k),
\qquad \rr_k\stackrel{\mathrm{iid}}{\sim}q_A .
\label{eq:expectation}
\end{equation}
Crucially $q_A$ is a Gaussian mixture that is sampled \emph{exactly}, with no
rejection: each atomic term is a scaled normal,
$o_i=(\pi/\alpha_i)^{3/2}\,\mathcal N(\rr;\cc_i,\tfrac{1}{2\alpha_i}\mathbf I)$, so
one draws a component $i$ with probability $\pi_i=(\pi/\alpha_i)^{3/2}/Z_A$ and
then $\rr\sim\mathcal N(\cc_i,\tfrac{1}{2\alpha_i}\mathbf I)$. Every sample lands
where $A$ has density-the variance reduction over uniformly darting a bounding
box, whose efficiency degrades with the box-to-molecule volume ratio.

Taking $f=m_B$ recovers the first-order overlap
$O^{\mathrm{FO}}_{AB}=Z_A\,\E_{q_A}[m_B]$, and the estimator
$\widehat O^{\mathrm{FO}}=\tfrac{Z_A}{n}\sum_k m_B(\rr_k)$ has variance
\begin{equation}
\operatorname{Var}\big[\widehat O^{\mathrm{FO}}\big]
=\frac{Z_A^{2}}{n}\operatorname{Var}_{q_A}[m_B]
=\frac{Z_A^{2}}{n}\Big(\E_{q_A}[m_B^{2}]-\E_{q_A}[m_B]^{2}\Big),
\label{eq:var}
\end{equation}
consistently estimated by the sample variance, so each run reports a standard
error $\mathrm{SE}=Z_A\,\widehat{\mathrm{std}}_{q_A}[m_B]/\sqrt n=O(n^{-1/2})$.
Two standard controls lower the constant: symmetrising the proposal over
$A\!\leftrightarrow\!B$ (averaging the estimates that sample from $q_A$ and from
$q_B$), and antithetic sampling-each Gaussian draw
$\rr=\cc_i+L_i\boldsymbol\varepsilon$ with $L_iL_i^{\top}=\tfrac{1}{2\alpha_i}\mathbf I$
is paired with its reflection $\cc_i-L_i\boldsymbol\varepsilon$, cancelling the
component of $m_B$ that is odd about $\cc_i$.

\subsection{Exact union overlap by bounded reweighting}
\label{sec:union}
The union occupancy~\eqref{eq:union} expands as the inclusion--exclusion series
\begin{equation}
u_A=\sum_{i\in A}o_i-\sum_{i<j}o_io_j+\sum_{i<j<k}o_io_jo_k-\cdots,
\label{eq:incexc}
\end{equation}
whose first term alone is the first-order model $m_A=\sum_i o_i$; the remaining
alternating terms are precisely the multiple-overlap corrections that first order
discards. Summing the series in closed form is combinatorial (up to $2^{N}-1$
terms), which is why production tools truncate it. Estimating the content of such
a union is, however, exactly the union-of-events setting in which importance
sampling is effective.\cite{owen2019importance} We therefore keep the samplable
mixture $q_A=m_A/Z_A$ as the proposal and correct it with the \emph{bounded}
weight $w_A(\rr)=u_A(\rr)/m_A(\rr)$, which satisfies $0\le w_A\le 1$ because
$0\le u_A\le m_A$ pointwise (a sum dominates one-minus-the-product of the same
occupancies). Then
\begin{equation}
O^{\cup}_{AB}=\int u_A u_B\,d\rr
=\int m_A\,\frac{u_A}{m_A}\,u_B\,d\rr
= Z_A\,\E_{\rr\sim q_A}\!\big[w_A(\rr)\,u_B(\rr)\big]
\approx \frac{Z_A}{n}\sum_{k=1}^{n} w_A(\rr_k)\,u_B(\rr_k).
\label{eq:union-is}
\end{equation}
This estimator is unbiased for the \emph{exact} union of all orders and evaluates
$u_A,u_B$ directly from Eqs.~\eqref{eq:occ}--\eqref{eq:union} at $O(N)$ cost per
sample, with no combinatorial enumeration. Its integrand is bounded,
$0\le w_A u_B\le 1$, so the variance is finite and controlled,
$\operatorname{Var}[\widehat O^{\cup}]\le Z_A^{2}/(4n)=O(n^{-1})$; a dense grid
quadrature of Eq.~\eqref{eq:union-overlap} provides an independent ground truth.

\subsection{Differentiable alignment and the hybrid pipeline}
\label{sec:align}
Both Eqs.~\eqref{eq:fo} and~\eqref{eq:union-is} are smooth in the pose of $B$. We
parameterise that pose by $\xi=(\boldsymbol\omega,\mathbf t)\in\mathbb R^6$ acting
rigidly on every centre of $B$,
\begin{equation}
\cc_j\;\longmapsto\;R(\boldsymbol\omega)\,\cc_j+\mathbf t,
\qquad R(\boldsymbol\omega)=\exp\big([\boldsymbol\omega]_\times\big)\in SO(3),
\label{eq:pose}
\end{equation}
with the matrix exponential of the skew form of $\boldsymbol\omega$ (Rodrigues'
formula). Drawing the sample points $\{\rr_k\}$ once from the fixed reference $A$
and \emph{holding them constant} through the optimisation (common random numbers),
the objective $\widehat O(\xi)=\tfrac{Z_A}{n}\sum_k w_A(\rr_k)\,u_B(\rr_k;\xi)$
depends on $\xi$ only through the moved molecule, so differentiating under the sum
gives the pathwise (reparameterised) gradient\cite{mohamed2020mcgrad}
\begin{equation}
\nabla_\xi\widehat O(\xi)=\frac{Z_A}{n}\sum_{k=1}^{n} w_A(\rr_k)\,\nabla_\xi u_B(\rr_k;\xi),
\label{eq:grad}
\end{equation}
which reuses the same points for value and gradient and is therefore low-variance;
in practice it is obtained by automatic differentiation. We maximise the overlap
by Adam\cite{kingma2015adam} from multiple starts-the identity, the four proper
principal-axis (inertial) alignments, and random rotations-and keep the best,
which makes recovery robust to the objective's multi-modality. Because the
first-order overlap~\eqref{eq:fo} is exact and cheap, we adopt a \emph{hybrid}
pipeline: align on $O^{\mathrm{FO}}$ (fast, deterministic, differentiable), then
\emph{score} the superposed pair with the accurate union
estimator~\eqref{eq:union-is}.

\subsection{Confidence-bounded screening}
\label{sec:screen}
Because each estimate carries a standard error, samples can be allocated
adaptively. For a query/candidate pair we draw batches until the delta-method
standard error of the Tanimoto,
$\mathrm{SE}(T)=\big|(O_{AA}+O_{BB})/(O_{AA}+O_{BB}-O_{AB})^2\big|\,\mathrm{SE}(O_{AB})$,
falls below a tolerance $\tau$ (or a budget cap is hit). Candidates that clearly
match or clearly miss reach $\tau$ in a few thousand samples; only borderline
pairs cost the full budget. This is the fixed-confidence stopping rule of
best-arm identification\cite{audibert2010bai} specialised to shape screening.

\subsection{Implementation}
\label{sec:impl}
All numerical kernels are pure JAX\cite{jax2018}: they \texttt{jit}-compile to a
single fused kernel and run unchanged on CPU/GPU/TPU, and the shape container is a
JAX pytree so \texttt{vmap}/\texttt{grad} apply directly. Molecules are read with
RDKit; conformers use ETKDGv3\cite{riniker2015etkdg} with MMFF optimisation, and
van der Waals radii come from RDKit's periodic table. The library
(\texttt{shape3d}) exposes the analytic overlap, the IS and union estimators, the
grid reference, alignment, and screening. Experiments use deterministic PRNG keys.

\section{Results and Discussion}

\subsection{Accuracy: the union estimator is unbiased; first order over-counts}
\label{sec:acc}
\begin{figure}[t]
\centering
\includegraphics[width=\linewidth]{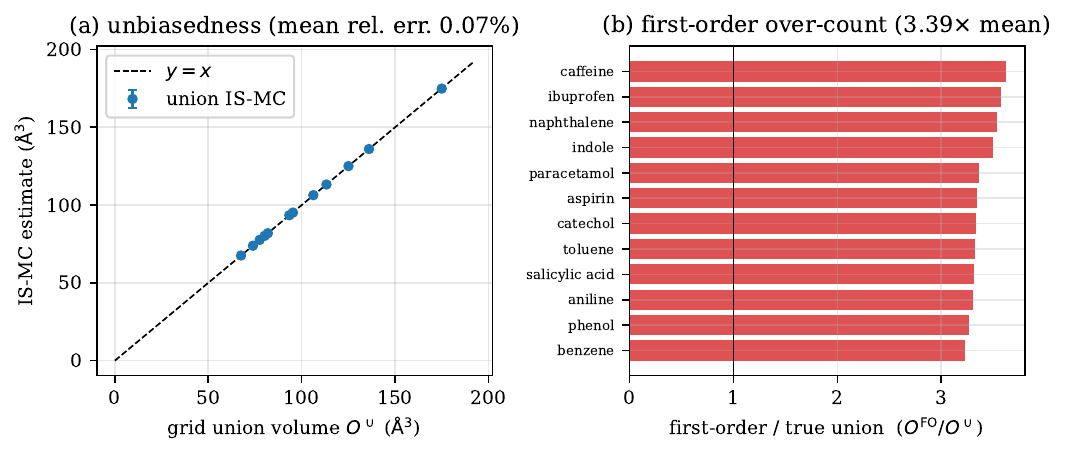}
\caption{(a) Union importance-sampling estimates versus grid ground truth for the
self-overlap of twelve molecules (mean relative error $0.07\%$). (b) The
first-order overlap over-counts the true union volume by $3.2$--$3.6\times$.}
\label{fig:accuracy}
\end{figure}

Figure~\ref{fig:accuracy} compares, over twelve drug-like molecules, the union
importance-sampling estimator~\eqref{eq:union-is} against grid quadrature and the
first-order analytic overlap~\eqref{eq:fo}. The IS estimates fall on the
$y=x$ line against grid truth with a mean relative error of $0.07\%$ (maximum
$0.16\%$ at $2\times10^5$ samples), confirming unbiasedness. In contrast the
first-order overlap over-counts the true union volume by a factor of $3.4$ on
average (range $3.2$--$3.6$). The first-order distortion is thus large and
systematic; it cancels partially in a Tanimoto ratio-which is why ROCS remains
useful-but it is not benign when absolute volumes or cross-model comparisons
matter, and it changes the fine-grained ranking relative to the exact union.

\subsection{Convergence}
\begin{figure}[t]
\centering
\includegraphics[width=\linewidth]{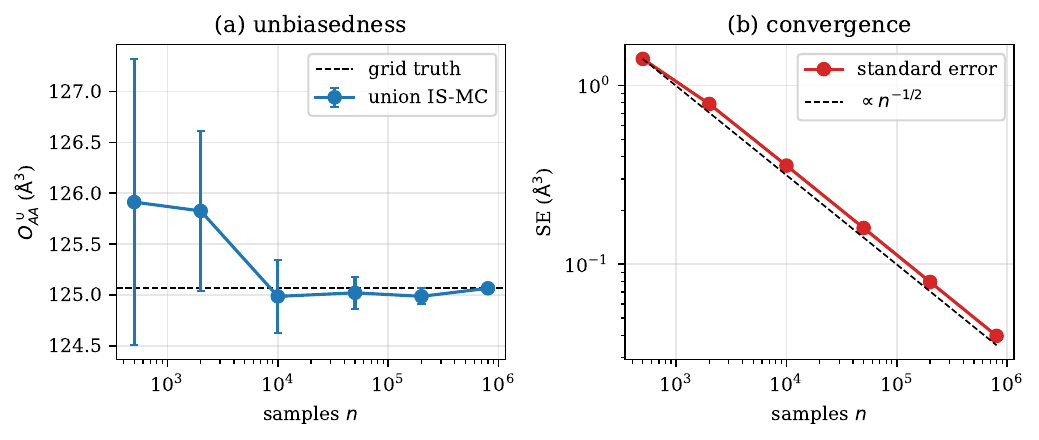}
\caption{Convergence of the union estimator (aspirin self-overlap). (a)
Unbiasedness across sample counts; (b) standard error $\propto n^{-1/2}$.}
\label{fig:conv}
\end{figure}

The estimator converges at the Monte-Carlo rate. Figure~\ref{fig:conv} shows the
union self-overlap of aspirin (2-acetoxybenzoic acid) as a function of sample
count: the estimate is
unbiased at every $n$ and the standard error decays as $n^{-1/2}$, reaching
$0.04~\Ang^3$ (relative error $<0.05\%$) at $8\times10^5$ samples against a grid
value of $125.1~\Ang^3$.

\subsection{Confidence-bounded screening}
\label{sec:screen-res}
\begin{figure}[t]
\centering
\includegraphics[width=\linewidth]{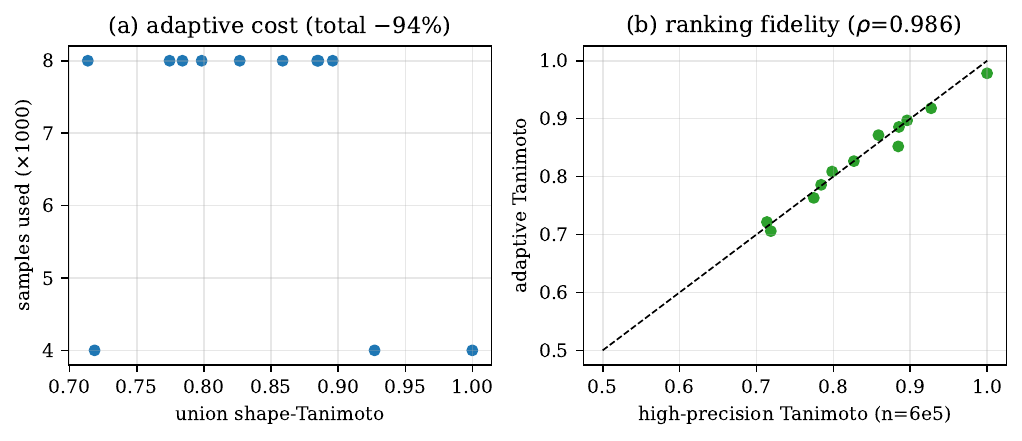}
\caption{Confidence-bounded screening of a 12-molecule library against aspirin.
(a) Adaptive sample cost per candidate versus its shape-Tanimoto; total sampling
is $94\%$ below a fixed budget. (b) Adaptive versus high-precision Tanimoto:
ranking is preserved (Spearman $\rho=0.99$).}
\label{fig:screen}
\end{figure}

Figure~\ref{fig:screen} screens a twelve-molecule library against an aspirin query
with the hybrid pipeline (first-order alignment, union scoring) under the adaptive
stopping rule ($\tau=0.01$). Sample cost tracks statistical difficulty: clear
matches and clear non-matches stop early, and the total sample budget is $94\%$
smaller than a fixed $1.2\times10^5$-per-candidate sweep ($8.4\times10^4$ versus
$1.44\times10^6$ samples). The adaptive ranking reproduces the high-precision
($6\times10^5$-sample) ranking with Spearman $\rho=0.99$. Because scoring uses the
exact union rather than first order, the ordering differs from a first-order
screen-a direct consequence of the $3.4\times$ over-count being molecule-dependent.

\subsection{Alignment}
\begin{figure}[t]
\centering
\includegraphics[width=\linewidth]{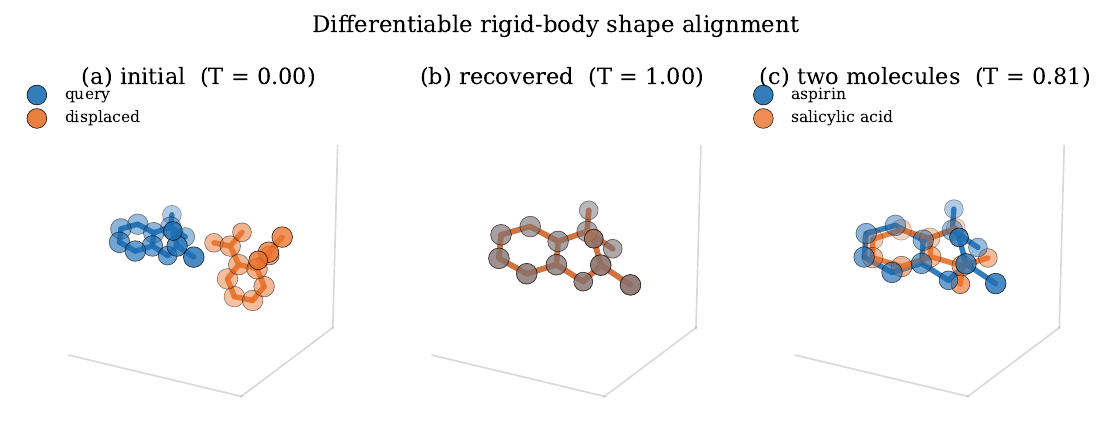}
\caption{Differentiable rigid-body shape alignment, shown as heavy-atom
stick-and-ball models with atoms sized by van der Waals radius. (a,b) Known-pose
recovery: a rigidly displaced copy of aspirin (orange), initially disjoint from
the query (blue, shape-Tanimoto $T=0.00$), is superposed exactly onto it
($T=1.00$; the two structures coincide). (c) Cross-molecule superposition of
salicylic acid (orange) onto aspirin (blue), $T=0.81$: the shared benzoic-acid
core registers while the substituents differ.}
\label{fig:align3d}
\end{figure}

\begin{figure}[t]
\centering
\includegraphics[width=\linewidth]{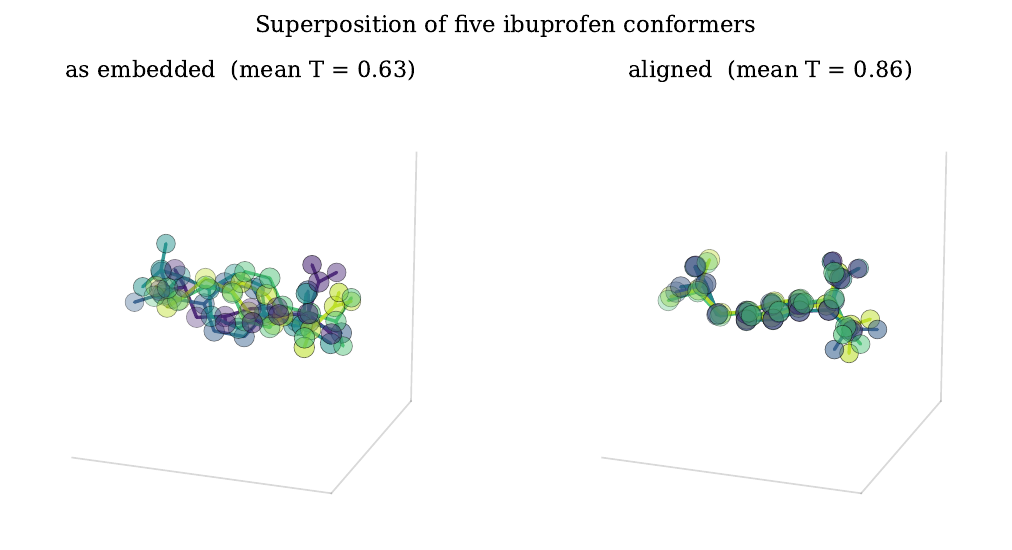}
\caption{Superposition of five ibuprofen conformers, coloured by conformer. As
embedded (left) the conformers form a diffuse cloud (mean pairwise shape-Tanimoto
$0.63$); after differentiable alignment (right) they register into a common
molecular frame (mean $0.86$), the residual reflecting real differences in
conformational shape. The same numbers underlie the pairwise matrix in
Figure~S1.}
\label{fig:conf3d}
\end{figure}

The differentiable multi-start optimiser recovers superpositions reliably
(Figure~\ref{fig:align3d}). A rigidly displaced copy of aspirin, disjoint from the
query at shape-Tanimoto $0.00$, is recovered to Tanimoto $1.00$
(Figure~\ref{fig:align3d}a,b); superposing two \emph{distinct} molecules-%
salicylic acid (2-hydroxybenzoic acid) onto aspirin-raises their shape-Tanimoto
from $0.63$ to $0.81$ (Figure~\ref{fig:align3d}c). The same machinery aligns
different conformers of a single flexible molecule: five conformers of ibuprofen
(2-[4-(2-methylpropyl)phenyl]propanoic acid) superpose from a mean cross-conformer
shape-Tanimoto of $0.63$ (as embedded) to $0.86$ after alignment
(Figure~\ref{fig:conf3d}), the residual below unity reflecting genuine
conformational shape differences; rigid benzene conformers are congruent and align
to $1.00$. The per-start score distributions, which reveal the multi-modality that
motivates multiple starts, are given in the Supporting Information.

\subsection{Retrospective virtual screening on DUD-E and LIT-PCBA}
\label{sec:bench}
To establish screening utility we ran a retrospective evaluation on the complete
DUD-E\cite{mysinger2012dude} (102 targets, $\sim$22.8k actives / 1.4M decoys) and
LIT-PCBA\cite{tran2020litpcba} (15 targets) benchmarks. For each target the
co-crystallised ligand serves as the 3D shape query; every library molecule is
embedded to a single ETKDGv3 conformer, rigidly aligned on the first-order
overlap, and scored by shape-Tanimoto. Enrichment is quantified with AUROC, the
enrichment factor (EF) and BEDROC\cite{truchon2007bedroc} using validated
implementations; the full protocol and per-target results are in the Supporting
Information (actives, and up to 1000 decoys per target, were used).

\begin{figure}[t]
\centering
\includegraphics[width=\linewidth]{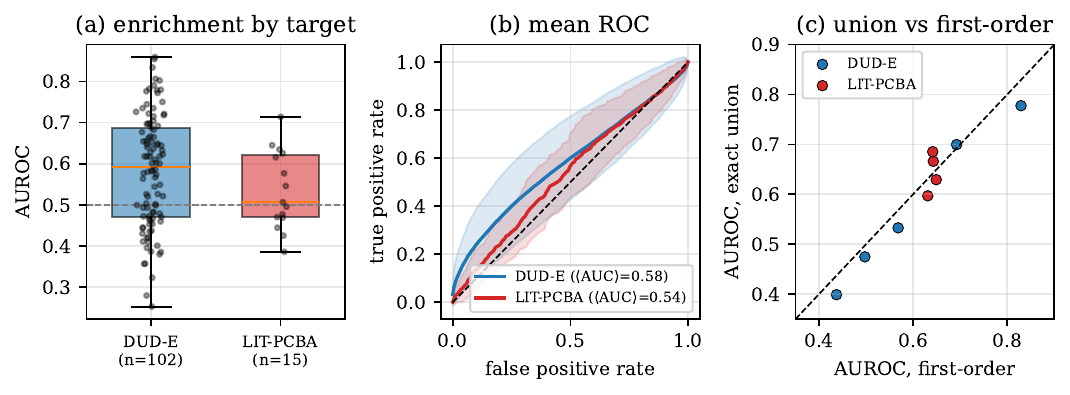}
\caption{Retrospective virtual screening. (a) Distribution of per-target AUROC on
the full DUD-E (102 targets) and LIT-PCBA (15 targets) panels; dashed line is
random. (b) Mean ROC curve per benchmark (band: $\pm1$ s.d.\ across targets). (c)
Per-target AUROC of the exact union estimator versus the first-order score;
points lie on $y=x$, i.e.\ the two give equivalent enrichment.}
\label{fig:bench}
\end{figure}

Figure~\ref{fig:bench} summarises the outcome. On DUD-E the mean AUROC is $0.58$
(median $0.59$; 21 of 102 targets above $0.70$, mean EF$_{1\%}=4.1$), and on
LIT-PCBA-built to be harder and less biased-the mean AUROC is $0.54$
(EF$_{1\%}=2.2$). These values are characteristic of single-conformer, shape-only
screening driven by one crystal-ligand query (cf.\ ROCS $\approx0.6$ on
DUD-E\cite{hawkins2007rocs}) and confirm that the estimator recovers genuine
shape-driven enrichment rather than a database artefact; per-target performance
spans easy shape targets (e.g.\ \emph{pyrd}, \emph{wee1}, \emph{fabp4}, AUROC
$0.83$--$0.86$) and shape-insensitive ones where a single query is unrepresentative.
Replacing the first-order score by the exact union estimator~\eqref{eq:union-is}
leaves aggregate enrichment statistically unchanged (paired
$\Delta\mathrm{AUROC}=-0.01$ over nine targets, Fig.~\ref{fig:bench}c): although
the union re-orders individual candidates relative to first order, the
shape-Tanimoto \emph{ratio} cancels the molecule-independent part of the
over-count, so ranking power is preserved. The
union estimator thus delivers the same enrichment as first order while
additionally providing the unbiased absolute volumes and per-estimate error bars
that make confidence-bounded screening possible. Conformer ensembles
(max-over-conformers), the standard accuracy lever for shape methods, are
supported by the library and are the natural route to higher enrichment.

\subsection{Relation to prior work, novelty, and limitations}
The analytic components-the Gaussian model,\cite{grant1995gaussian,grant1996fast}
first-order overlap and shape Tanimoto,\cite{hawkins2007rocs} and multi-start
inertial alignment-are standard, and WEGA\cite{yan2013wega} already corrects
first-order over-counting by per-atom reweighting. Our contribution is orthogonal
and, to our knowledge, new to shape-based VS: a \emph{stochastic} overlap
estimator that (i) importance-samples from the molecule's own mixture, (ii) reaches
the exact union volume by bounded reweighting, and (iii) exposes a standard error
that we turn into adaptive, confidence-bounded screening. The probability-product-kernel
view\cite{jebara2004ppk} places the estimator on firm statistical ground, and the
JAX implementation makes it differentiable and accelerator-portable, complementing
GPU analytic engines.\cite{haque2010paper,roshambo2_2025}

Two limitations frame future work. First, for the \emph{first-order} quantity the
analytic form is exact and cheaper than Monte Carlo; the stochastic estimator pays
off specifically for the exact union (intractable analytically at scale) and for
its error-bar-driven adaptivity. Second, the retrospective evaluation
above uses a single conformer and a single crystal-ligand
query per target; conformer ensembles and a head-to-head comparison against
production ROCS and ROSHAMBO\cite{atwi2024roshambo,roshambo2_2025} engines under a
common protocol are the natural next steps toward a definitive screening
benchmark.

\section{Conclusions}
We have introduced an importance-sampling estimator of Gaussian molecular shape
overlap-to our knowledge the first stochastic treatment of this quantity-that
is unbiased, carries an analytic standard error, and, through a bounded importance
weight, estimates the exact union volume of all inclusion--exclusion orders at
$O(N)$ cost per sample. It removes the $3.4\times$ first-order over-count
(matching grid quadrature to $0.07\%$), enables confidence-bounded screening that
saves $94\%$ of samples at fixed ranking fidelity, and, being differentiable,
drives robust gradient-based alignment. The approach is released as an
open-source, GPU-portable JAX library and offers a statistically principled,
error-aware complement to deterministic analytic shape screening.

\begin{suppinfo}
Derivation of the unbiasedness and variance of the estimators; grid-convergence
study; alignment and conformer-overlay experiments; software and reproducibility
details (PDF).
\end{suppinfo}

\bibliography{references}

\end{document}